\documentclass[11pt]{article}
\pdfoutput=1
\usepackage{jheppub}
\usepackage{epsfig}
\usepackage{amsfonts}
\usepackage{amsmath}
\usepackage{amssymb}
\usepackage{bbm,bm}
\usepackage{slashed}
\usepackage{hyperref}

\usepackage{tikz}
\newdimen\nodeDist
\allowdisplaybreaks[1]

\def\Eq#1{Eq.~(\ref{#1})}
\def\OO{\mathcal{O}}

\def\alphas{\alpha_{\mathrm{s}}}

\def\Tr{\,\mathrm{Tr}\:}
\def\Det{\,\mathrm{Det}\:}
\def\Dslash{\slashed{\mathrm{D}}}

\def\PPQ{P_{\!_{\mathrm{PQ}}}}
\def\ZPQ{Z_{_{\mathrm{PQ}}}}

\title{Bounding the QCD Equation of State with the Lattice}
\author[a]{Guy D.\ Moore}
\author[a,b]{Tyler Gorda}
\affiliation[a]{
Institut f\"ur Kernphysik, Technische Universit\"at Darmstadt,
Schlossgartenstrasse 2,\\
D-64289 Darmstadt, Germany}
\affiliation[b]{ExtreMe Matter Institute EMMI, GSI Helmholtzzentrum f\"ur Schwerionenforschung GmbH, 64291 Darmstadt, Germany}
\emailAdd{guy.moore@physik.tu-darmstadt.de}
\emailAdd{gorda@itp.uni-frankfurt.de}

\abstract{
The equation of state of QCD matter at high densities is relevant for neutron star structure and for neutron star mergers and has been a focus of recent work.
We show how lattice QCD simulations, free of sign problems, can provide an \textsl{upper bound} on the pressure as a function of quark chemical potentials.
We show that at large chemical potentials this bound should become quite sharp;
the difference between the upper bound on the pressure $\PPQ$ and the true pressure $P$ is of order $\PPQ-P = \OO(\alphas^3 P)$.
The corrections arise from a single Feynman diagram; its calculation would render remaining corrections $\OO(\alphas^4 P)$.
}


\begin{document}
\maketitle

%
\section{Introduction}
\label{sec:intro}

The equation of state (EOS) of strongly interacting matter dictates the thermodynamics of any system ultimately composed of quarks and gluons. At high temperatures and low net baryon densities, the EOS can be computed directly from the partition function of Quantum Chromodynamics (QCD) using Monte-Carlo lattice techniques
\cite{Borsanyi:2013bia,HotQCD:2014kol}
and compared to experimental determinations of thermodynamic properties \cite{Gardim:2019xjs}.
However, at low temperatures and high net baryon densities, such techniques fail due to the well-known sign problem \cite{deForcrand:2009zkb,Nagata:2021ugx}, and alternative methods are required to determine the EOS.  

Such cold and dense QCD matter is in fact realized in nature in the cores of massive neutron stars, where matter is in equilibrium under the weak interactions (beta equilibrated) and dense enough to populate the three lightest quark flavors. In this context, the EOS is intimately tied to the bulk properties of neutron stars via the stellar structure equations of general relativity \cite{Tolman:1939jz,Oppenheimer:1939ne}, and hence observations of neutron star properties, such as masses \cite{Antoniadis:2013pzd,Cromartie:2019kug,Fonseca:2021wxt}, tidal deformabilities \cite{TheLIGOScientific:2017qsa,LIGOScientific:2018cki,LIGOScientific:2018hze}, and radii \cite{Steiner:2017vmg,Nattila:2017wtj,Shawn:2018,Miller:2019cac,Riley:2019yda,Miller:2021qha,Riley:2021pdl} can be used to constrain the EOS of QCD. 
Recently the community has converged on a strategy for inferring the EOS of cold and dense QCD matter using these astrophysical constraints in combination with a general set of causal extensions of low-density effective-field-theory calculations \cite{Tews:2012fj,Lynn:2015jua,Drischler:2017wtt,Drischler:2020hwi,Keller:2022crb} of the EOS of dense nuclear matter (see, e.g., \cite{Hebeler:2013nza, Tews:2018iwm,Landry:2018prl,Dietrich:2020efo,Capano:2019eae,Raaijmakers:2019dks,Landry:2020vaw,Essick:2019ldf,Miller:2019nzo,Al-Mamun:2020vzu,Miller:2021qha,Essick:2021kjb,Huth:2021bsp,Lim:2022fap,Essick:2023fso}), sometimes also paired with constraints from perturbative QCD calculations at high net baryon densities \cite{Kurkela:2009gj,Kurkela:2016was,Gorda:2018gpy,Gorda:2021znl,Komoltsev:2021jzg,Gorda:2021gha,Gorda:2023mkk,Gorda:2023usm} (see, e.g.  \cite{Kurkela:2014vha,Annala:2017llu,Most:2018hfd,Altiparmak:2022bke,Annala:2019puf,Annala:2021gom,Gorda:2022jvk,Somasundaram:2022ztm,Han:2022rug, Jiang:2022tps,
Annala:2023cwx,Brandes:2023hma,Mroczek:2023zxo}). In addition to these constraints, there has been recent work on incorporating other experimental constraints, such as bounds from low-energy nuclear collision experiments \cite{Huth:2021bsp,Sorensen:2023zkk}, which have so far been seen to compliment those from astrophysics. On the theoretical side, the unitary gas constraint \cite{2002PThPS.146..363C,Schwenk:2005ka,Carlson:2012mh,Gandolfi:2015jma,Tews:2016jhi} -- which has been conjectured to provide a lower bound of the energy per particle at low to moderate net baryon densities in the hadronic phase -- has been used as a reference to benchmark different nuclear-theory calculations of the hadronic EOS as they are extended to higher densities.

In this paper we argue that one more constraint (or family of constraints) can be added to this list of bounds on the dense QCD EOS.
While lattice QCD cannot compute the EOS for the physical combination of chemical potentials, \textsl{phase-quenched} lattice QCD is free of sign problems.
The pressure as a function of chemical potentials, calculated using phase quenching, $\PPQ(\mu_q)$, is a \textsl{strict upper bound} on the true pressure at the same values of quark chemical potentials: 
$\PPQ(\mu_q) \geq P(\mu_q)$.
This fact was established by Cohen in Ref.~\cite{Cohen:2003ut}, who used it to show how the case of an isospin chemical potential -- opposite up and down quark chemical potentials -- can be used to bound the more physically interesting case of equal up and down quark chemical potentials.
Unfortunately, the combination of chemical potentials relevant for neutron stars is rather different: the down and strange chemical potentials should be equal $\mu_d = \mu_s$, while the up-quark chemical potential is somewhat smaller to accommodate charge neutrality, $\mu_u < \mu_d$.
Bounding the equation of state for this combination of chemical potentials starting from lattice results for isospin chemical potentials
requires the use of additional inequalities
\cite{Lee:2004hc},
as recently explored by Fujimoto and Reddy \cite{Fujimoto:2023unl}.

We argue here that the most effective way to use the lattice to bound the neutron-star equation of state is to perform new phase-quenched lattice simulations using the physical combination of up, down, and strange quark chemical potentials.
Unlike an isospin chemical potential, the phase-quenched version of such a combination represents a completely unphysical system.
But we expect that it will present the tightest bounds on the physical equation of state.
In particular, we show here that, while at low densities the constraint is likely to be very loose, at high densities it should become ever sharper.
In fact, we will demonstrate that, in the perturbative region, the relative difference  $\frac{\PPQ(\mu_q) - P(\mu_q)}{P(\mu_q)}$ is of order $\alphas^3$.
Furthermore, at this order in the coupling expansion, the difference arises from a \textsl{single} Feynman diagram, which we identify.%
\footnote{Two diagrams if one considers the two directions which  fermion arrows in a loop can point to represent distinct diagrams.}

An outline of this paper is as follows.
In the next section we review the path integral for QCD at finite chemical potential, and how it is related to the phase-quenched version.
The section reviews the (quite simple) proof that $\PPQ(\mu_q) \geq P(\mu_q)$, and discusses a little more how one should interpret the phase-quenched calculation.
In Section \ref{sec:pert} we show how to represent the (unphysical) phase-quenched theory in Feynman diagrams, and we identify the unique $\OO(\alphas^3)$ diagram which differs between $\PPQ(\mu_q)$ and $P(\mu_q)$.
Section \ref{sec:latt} estimates the size of lattice artifacts in evaluating the pressure on the lattice, in order to give guidance for how small the lattice spacing must be at a given, large $\mu$ value.
We end with a discussion, which lists the most important directions for further work.

\section{Path integral and phase quenching}
\label{sec:overview}

Here we review the proof that the pressure from the phase-quenched theory is a strict upper bound on the original theory.
The proof as we formulate it is due to Cohen \cite{Cohen:2003ut}, though the underlying ideas are older \cite{Alford:1998sd}.
The partition function of QCD at a small finite temperature $T = 1/\beta$ in a box with a large volume $V$ and with quark chemical potentials $\mu_q$ is%
\footnote{The integral as written requires either gauge fixing or a restriction that $\mathcal{D} G_\mu$ should be understood only to run over distinct affine connections, but this point is not relevant here, since the fermionic determinants are our main focus.}
\begin{equation}
\label{Zdef}
    Z(\beta,\mu_q) = \int \mathcal{D}G_\mu \,
    \exp \left(-\int_0^\beta dt \int_V d^3 x \; \mathcal{L}_{\mathrm{E}}(G) \right)
    \; \prod_{q=u,d,s} \Det \Big( \Dslash +m_q + \mu_q \gamma^0 \Big) .
\end{equation}
Because $\Dslash+m_q$ and $\mu_q \gamma^0$ have different $\gamma^5$-Hermiticity \cite{deForcrand:2009zkb}, each determinant is complex.
Phase quenching is the replacement of $Z$ with
\begin{equation}
\label{ZPQdef}
    \ZPQ(\beta,\mu_q) \equiv \int \mathcal{D}G_\mu \,
    \exp \left(-\int_0^\beta dt \int_V d^3 x \; \mathcal{L}_{\mathrm{E}}(G) \right)
    \; \prod_{q=u,d,s} \left| \Det \Big( \Dslash +m_q + \mu_q \gamma^0 \Big) \right|.
\end{equation}
That is, one uses the absolute values of the determinants, rather than the determinants themselves.
Since the Euclidean gluonic action $\mathcal{L}_{\mathrm{E}}(G)$
is strictly real,%
\footnote{We assume that the QCD theta angle
\cite{tHooft:1976,Jackiw:1976pf,Callan:1979bg}
is zero.}
the integrand is now real and positive and equals the absolute value of the integrand for $Z$.
Since the integral of the absolute value of a function over a positive measure is greater than or equal to the integral of the original complex function, we have
\begin{equation}
\label{inequality}
\ZPQ(\beta,\mu_q) \geq Z(\beta,\mu_q) 
\quad \mbox{and therefore} \quad
\PPQ(\beta,\mu_q) \geq P(\beta,\mu_q) \,,
\end{equation}
where $P = \ln(Z)/(\beta V)$ is the pressure.%
\footnote{In practice we want $P(\mu_q) - P(\mu_q=0)$, that is, one should subtract the zero-chemical-potential value of the pressure.  This has the benefit of removing cosmological-constant-type power UV divergences.}

Let us investigate the interpretation of the phase-quenched theory.
Applying $\gamma^5$-Hermiticity, one finds that
\cite{Cohen:2003kd,deForcrand:2009zkb}
\begin{equation}
\label{eqconjugate}
    \left( \Det \Big( \Dslash + m_q + \mu_q \gamma^0 \Big) \right)^*
    = \left( \Det \Big( \Dslash + m_q - \mu_q \gamma^0 \Big) \right)
\end{equation}
and therefore
\begin{align}
\label{squareroot} 
    \left| \Det \Big( \Dslash + m_q + \mu_q \gamma^0 \Big) \Big) \right|
    & = \sqrt{
    \Det \Big( \Dslash + m_q + \mu_q \gamma^0 \Big)
    \Det\Big( \Dslash + m_q - \mu_q \gamma^0 \Big)   }\,,
    \\
    \label{ZPQ2}
    \ZPQ & = \int \mathcal{D}G_\mu \,
    \exp \left(-\int_0^\beta dt \int d^3 x \mathcal{L}_{\mathrm{E}}(G) \right)
    \\ \nonumber  & \qquad \times \prod_{q=u,d,s} \sqrt{
    \Det \Big( \Dslash +m_q + \mu_q \gamma^0 \Big) 
    \Det \Big( \Dslash +m_q - \mu_q \gamma^0 \Big) }.
\end{align}
Therefore, the phase-quenched theory is equivalent to a theory with twice as many fermionic species, but where each is represented by the square root of a determinant -- 
think of each as a half-species, which appear in pairs with equal mass but opposite chemical potential.
This theory is clearly unphysical, since a half-species of fermion does not make sense as an external state.
A well known exception is if two quark masses and chemical potentials are equal
\cite{Alford:1998sd,Son:2000xc,Son:2000by,Splittorff:2000mm}.
In particular, for $m_u = m_d$ and $\mu_u = \mu_d$ with $\mu_s = 0$,
representing ``standard'' baryonic chemical potential,
the phase-quenched version is equivalent, after re-labeling two of the half-species,
to the case $\mu_d = -\mu_u$, corresponding to an isospin chemical potential.
This among other things has motivated rather detailed lattice investigations of the case of finite isospin chemical potential, see for instance 
Refs.~\cite{Kogut:2002tm,Kogut:2002zg,Nishida:2003fb,Kogut:2004zg,Beane:2007es,Detmold:2008fn,Detmold:2012wc,Endrodi:2014lja,Janssen:2015lda,Brandt:2017oyy,Brandt:2019hel,Brandt:2021yhc,Brandt:2022hwy,Abbott:2023coj} or the comprehensive review by Mannarelli \cite{Mannarelli:2019hgn}.
In this case it is particularly clear that, for small chemical potentials, the two pressures are quite different.
Tom Cohen has analyzed the case for an isospin chemical potential
\cite{Cohen:2003kd}
and shown that nonzero $\mu_q$ in \Eq{squareroot} does not change the value of the determinant until $\mu_q = m_\pi/2$, when eigenvalues of $\Dslash+m_q+\mu_q \gamma_0$ start to cross the imaginary axis.
He refers to the constancy of the determinant up to this point as ``the silver blaze problem.''
Applying his analysis for the general case of distinct $\mu_u,\mu_d,\mu_s$, one finds that $\PPQ$ first deviates from its vacuum value when either $\mu_d = m_\pi/2$, $\mu_u=m_\pi/2$, or $\mu_s = m_{s\bar{s}}/2$, where $m_{s\bar{s}}$ is the mass of the unphysical strange-antistrange pseudoscalar obtained by ignoring disconnected contributions to the correlation function.
Therefore, at small chemical potentials $m_\pi/2 < \mu_q < m_p/3$,
$\PPQ$ differs significantly from the physical pressure.

In the next section we will see that, for large chemical potentials, the two pressures become much more similar.

\section{Perturbation theory}
\label{sec:pert}

To construct a perturbation theory for \Eq{ZPQ2}, as usual for fermions, we rewrite
\begin{align}
    \Det(A) & = \exp( \ln\Det (A) ) = \exp(\Tr \ln(A))
    \nonumber  \\
    \sqrt{\Det (A)} & = \exp\left( \frac 12 \ln \Det(A) \right)
    = \exp \left( \frac 12 \Tr \ln(A) \right) \,.
\label{tracelog}
\end{align}
The process of deriving Feynman rules from $\frac 12 \Tr \ln(\Dslash+m\pm \mu \gamma^0)$ is the same as without the $\frac 12$ factor, \textsl{except} that each fermionic loop receives a factor of $\frac 12$.
Therefore, in each place where a fermionic loop can appear, instead of performing a sum over $n_f$ fermions, each with mass $m_q$ and chemical potential $\mu_q$, we sum over $2n_f$ terms; twice for each $m_q$, once with $\mu_q$ and once with $-\mu_q$ but each with an overall factor of $\frac 12$.
This is the same as averaging the loop over whether $\mu$ is positive or negative.

To see the impact of $\mu \to -\mu$ we remind the reader about Furry's theorem
\cite{Furry:1937zz}.
Consider a fermion loop with $n$ external gluons connected by $n$ propagators, with incoming momenta $q_1,\ldots,q_n$ with
$q_n = -\sum_{j=1}^{n-1} q_j$ by momentum conservation.
The fermions are in a representation $R$ (typically the fundamental representation) with Hermitian generators $T^A$.
Its group-theory factor and the numerator of its Feynman rule are
\begin{align}
\label{furry1}
    \Tr \left( T^{A_1} \ldots T^{A_n}  \right) \;
    \Tr \gamma_{\mu_1} (i\slashed{p}_1+m+\mu \gamma^0)
    \gamma_{\mu_2} (i \slashed{p}_2+m + \mu \gamma^0) \ldots
\end{align}
where $p_1$ is the loop momentum and $p_i = p_{i-1} + q_i$.
Inserting the charge conjugation matrix times its inverse, $C C^{-1}$, between each neighboring term and using 
$C^{-1} \gamma_\mu C = - \gamma_\mu^\top$, we find
\begin{align}
\label{furry2}
    \Tr \left( (-T^{A_1})\ldots (-T^{A_n}) \right) \;
    \Tr \gamma_{\mu_1}^\top (-i \slashed{p}^\top_1 + m - \mu \gamma^{0\top}) \gamma_{\mu_2}^\top 
    (-i \slashed{p}^\top_2 + m - \mu \gamma^{0\top}) \ldots  .  
\end{align}
Reversing the order of the symbols to get rid of the transposes on the gamma matrices, and using $T^\top = T^*$ for the Hermitian group generators, we find
\begin{align}
\label{furry3}
    \Tr \left( (-T^{*A_n}) \ldots (-T^{*A_1}) \right)
    \Tr \ldots (-i\slashed{p}_2 + m - \mu \gamma^0) \gamma_{\mu_2}
    (-i \slashed{p}_i + m -\mu \gamma^0) \gamma_{\mu_1}
\end{align}
which is the same diagram,%
\footnote{Reversing the sense in which the loop is traversed flips the signs of the momenta as well as reversing the order of the matrices in the color and Dirac traces.}
traversed in the opposite sense, with $\mu \to -\mu$ and with $T^A \to (-T^{*A})$ which is the generator of the conjugate representation -- so if the $T^A$ generate the fundamental representation, the $-T^{*A}$ generate the antifundamental representation.
Another way to state Furry's theorem is then that reversing the sign of $\mu$ is equivalent to considering fermions with the original $\mu$ value but in the conjugate representation (quarks with $-\mu$ are the same as antiquarks with $+\mu$).

For the case of two external gluons we have
\begin{equation}
\label{trace2}
    2 \Tr T^A T^B = \delta^{AB} = 2 \Tr (-T^{*A})(-T^{*B})
\end{equation}
and the fundamental and antifundamental representations give the same answer.
The representations first differ when there are three gluons attached:
\begin{align}
\label{trace3}
    4 \Tr T^A T^B T^C & = i f^{ABC}
    + d^{ABC} & \mbox{but}
    \nonumber \\
    4 \Tr (-T^{*A})(-T^{*B})(-T^{*C}) & = i f^{ABC} - d^{ABC}
\end{align}
where $d^{ABC}$ is a totally symmetric symbol which arises in groups of rank 2 or more.

\begin{figure}[tbh]
    \centerline{
    \includegraphics[width=0.8\textwidth]{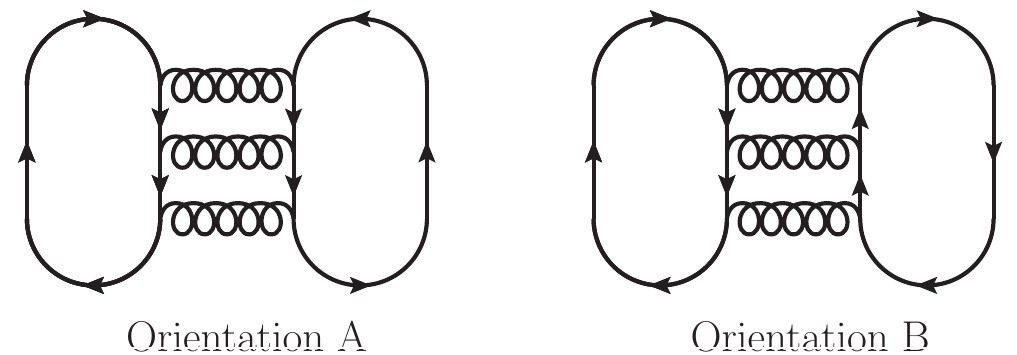}}
    \caption{The lowest-order diagram which distinguishes between fundamental and antifundamental representations, with two relative orientations for the fermionic loops.
    \label{fig:threegluon}
    }
\end{figure}

Because both $f^{ABC} d^{ABD} = 0$ and $\delta^{AB} d^{ABC} = 0$, the first group-theoretical structure where $d^{ABC}$ can give a nonzero contribution, for the group SU($N_c$), is
\begin{equation}
\label{dcontraction}
    d^{ABC} d^{ABD} = \frac{N_c^2-4}{N_c} \delta^{CD} \qquad
    \mbox{so} \qquad
    d^{ABC} d^{ABC} = \frac{(N_c^2-1)(N_c^2-4)}{N_c} .
\end{equation}
This contraction can only appear in a diagram with at least two fermion loops with at least three gluon attachments each.
The only such diagram at the $\alphas^3$ level is shown in Figure \ref{fig:threegluon}.
Fermion number can traverse the two loops with two relative orientations, shown on the left and the right in the figure.
We will write the two versions of the diagram, stripping off all group-theory factors, for two specific species $(i,j)$, as
$A(\mu_i,\mu_j)$ and $B(\mu_i,\mu_j)$.
That is, $A(\mu_i,\mu_j)$ and $B(\mu_i,\mu_j)$ represent the value of each fermion orientation in the abelian version of the diagram.
In this notation, the contribution of this diagram to the true pressure is
\begin{align}
\label{deltaP}
  \delta P =  & \left( \Tr T^A T^B T^C \: \Tr T^A T^B T^C \right) A(\mu_i,\mu_j)
    + \left( \Tr T^A T^B T^C \: \Tr T^C T^B T^A \right) B(\mu_i,\mu_j)
\nonumber \\
= &
\frac{(N_c^2-1)(N_c^2-4)}{16 N_c}\left( A(\mu_i,\mu_j) + B(\mu_i,\mu_j) \right)
+ 
\frac{(N_c^2-1) N_c}{16}\left( B(\mu_i,\mu_j) - A(\mu_i,\mu_j) \right)
.
\end{align}
In the abelian theory the group-theoretical coefficients on the first and second terms of the second line would be 1 and 0 respectively.

Furry's Theorem means that
$B(\mu_i,\mu_j) = - A(\mu_i,-\mu_j) = -A(-\mu_i,\mu_j)$.
This allows us to rewrite \Eq{deltaP} in terms of $A$ only:
\begin{equation}
    \label{deltaPA}
    \delta P = \frac{(N_c^2-1)(N_c^2-4)}{16 N_c}\left( A(\mu_i,\mu_j) - A(\mu_i,-\mu_j) \right)
-
\frac{(N_c^2-1) N_c}{16}\left( A(\mu_i,\mu_j) + A(\mu_i,-\mu_j) \right).
\end{equation}
Averaging over $\mu_j \leftrightarrow -\mu_j$ and/or $\mu_i \leftrightarrow -\mu_i$, as we are instructed to do in evaluating $\PPQ$, eliminates the first expression and leaves only the second.
Therefore, the difference between the phase-quenched and true pressure, at order $\alphas^3$, is
\begin{align}
\label{PPQminusP}
    \PPQ - P & = - \sum_{i,j=uds}
    \frac{(N_c^2-1)(N_c^2-4)}{16N_c} \left( A(\mu_i,\mu_j) - A(\mu_i,-\mu_j) \right).
\end{align}
We can show that this combination is automatically positive; the proof, along with an explicit evaluation, will appear in a forthcoming publication.

Evaluating this single combination of diagrams would determine the perturbative correction between the phase-quenched and true pressures, through to order $\alphas^3$.
Any quark with $\mu_i=0$ does not contribute to \Eq{PPQminusP}.
The contribution of soft gluon momenta to \Eq{PPQminusP} is suppressed, since the three-gluon hard loop
contains only the $f^{ABC}$ group-theory factor \cite{Braaten:1989mz}.\footnote{We are indebted to Saga S\"appi for useful conversations on this point.}
Therefore \Eq{PPQminusP} does not contain any logarithmically enhanced
$\sim \alphas^3 \ln(\alphas)$ contributions.

As an aside, we comment on the behavior of these terms as a function of $N_c$.
In the large $N_c$ or t'Hooft limit, the group-theory factor on Orientation $B$ in \Eq{deltaP} is $\propto N_c^3$ while that for Orientation $A$ is $\propto N_c$ and is suppressed.
In contrast, for the group SU(2), the group-theoretical factor on $A+B$ in \Eq{deltaP} vanishes and there is no distinction between the fundamental and antifundamental representations.
In fact, because every SU(2) representation is equivalent to its conjugate representation, it is easy to show that any combination of chemical potentials is free of sign problems in 2-color QCD, which has motivated investigation of this theory at finite density
\cite{Kogut:1999iv,Kogut:2000ek,Cotter:2012mb,Astrakhantsev:2020tdl,Boz:2019enj,Iida:2022hyy,Braguta:2023yhd}.

\section{Lattice considerations}
\label{sec:latt}

The previous sections have made clear that a lattice calculation of $\PPQ$ would be valuable.
Here we want to take one small step towards estimating how difficult such a lattice calculation would be.
The larger $\mu a$ is, the more rapidly a lattice calculation develops statistical power; since $P \propto \mu^4$, we expect that the signal-to-noise should scale roughly as $(\mu a)^4$.
Therefore it is advantageous to work on lattices with the largest $(\mu a)$ we can get away with.

But increasing $\mu a$ increases systematic lattice-spacing effects.
In general, for a lattice calculation to be accurate we need the lattice spacing to obey $a \Lambda_{\mathrm{QCD}} \ll 1$.
But $\mu$ introduces an additional scale, and for the regime $\mu \gg \Lambda_{\mathrm{QCD}}$, we expect to need the stronger condition $\mu a \ll 1$.
But what does $\mu a \ll 1$ really mean -- that is, how small does $\mu a$ really need to be?
To estimate this, we compute how different the vacuum-subtracted%
\footnote{The pressure receives divergent vacuum contributions (the cosmological constant problem) which have to be subtracted in a lattice treatment.}
pressure $P(\mu) - P(0)$ is on the lattice from its continuum value, at lowest (zero) order in the strong coupling.
Here we will only consider staggered quarks
\cite{Kogut:1974ag,Banks:1975gq,Susskind:1976jm},
since we expect this quark formulation to be used in practical calculations.

In the continuum, one free Dirac fermion with chemical potential $\mu$ provides a pressure of
\begin{align}
\label{Pversion1}
    P & = \frac{1}{\beta V} \ln(Z)
    = \int \frac{d^4 p}{(2\pi)^4} \left(
    \Tr \ln (-i \slashed{p}+m+\mu \gamma^0)  
    - \Tr \ln(-i \slashed{p}+m) \right) \,,
\end{align}
where the second term subtracts the $\mu=0$ ``vacuum'' pressure.
Evaluating the trace, one finds
\begin{align}
\label{Pversion2}
    P & = 2 \int \frac{d^3 \vec p}{(2\pi)^3}
    \int \frac{dp_0}{2\pi}
    \left( \ln \frac{(p_0+i\mu)^2 + \vec p^2 + m^2}{p_0^2 + \vec p^2 + m^2} \right).
\end{align}
Naively it appears that $\mu$ contributes to the pressure at any $\vec p$ value including when $\sqrt{\vec p^2 + m^2} > |\mu|$, but this is illusory.
Separating the numerator and denominator of the log, one may deform the $p^0$ integration contour for the former, shifting it by $-i\mu$.
If $\sqrt{p^2+m^2} > |\mu|$ then this contour deformation encounters no singularities, and the $p_0$ integral at nonzero and at zero $\mu$ are identical.
For $|\mu| > \sqrt{\vec p^2 + m^2}$ there is a cut running from
$p^0 = 0$ to $p^0 = -i(\mu- \sqrt{\vec p^2+m^2})$ with discontinuity $2\pi i$ which the deformed contour must enclose, leading to
\begin{align}
\label{Pversion3}
    P & = 2 \int \frac{d^3 \vec p}{(2\pi)^3}
    \left( |\mu| - \sqrt{\vec p^2 + m^2} \right)
    \Theta \left( |\mu| - \sqrt{\vec p^2+m^2} \right)
\end{align}
as expected.
The expressions for the energy density $\varepsilon$ and for $\mu N$ the chemical potential times the number density are the same but with the first factor, 
$|\mu| - \sqrt{\vec p^2+m^2}$,
replaced by $\sqrt{\vec p^2+m^2}$ and by $|\mu|$ respectively, recovering the thermodynamical relation $\mu N = \varepsilon + P$.

Naive staggered fermions
\cite{Susskind:1976jm}
represent four species of physical fermions with the limitation that each $p_\mu$ component runs over $[-\pi/(2a),\pi/(2a)]$ with $a$ the lattice spacing%
\footnote{The Brillouin zone should run over $[-\pi/a,\pi/a]$ but it is divided into 16 regions in which each component of $p$ runs over half as large a range, $p \in [-\pi/2a,\pi/2a]$.
These regions represent 16 ``doubler'' fermions, and the staggering procedure removes a factor of 4.  
For a review see \cite{Gattringer:2010zz}.}
and with the substitution
\begin{align}
\label{nearestneighbor}
    \partial_j \psi(x) \quad \Longrightarrow \quad &  
    \frac{\psi(x+a\hat{j}) - \psi(x-a \hat j)}{2a} \,,
    \nonumber \\
    \partial_0 \psi(x) + \mu \psi(x) \quad \Longrightarrow \quad &
    \frac{ e^{a\mu} \psi(x+a\hat{0}) - e^{-a\mu} \psi(x-a \hat{0})}{2a} \,,
    \nonumber \\
    -i p_j \quad \Longrightarrow \quad & 
    \frac{-i}{a} \sin (a p_j) \,,
    \nonumber \\
    -ip_0 + \mu \quad \Longrightarrow \quad & 
    \frac{1}{a} \left( -i\sin(a p_0) \cosh(a \mu) + \cos(a p_0) \sinh(a \mu) \right) \,.
\end{align}
Here $\hat{j}$ and $\hat{0}$ represent the unit vector in the $j$ spatial direction and in the time direction respectively.
Note that the effects of $\mu \gamma^0$ must be incorporated along with the time derivative because, in the staggered formulation, an insertion of $\gamma^0$ requires that the comparison be made between $\bar\psi$ and $\psi$ which differ by an odd number of steps in the time direction.
This is in any case the preferred way of introducing a chemical potential because it avoids quadratic-in-$\mu$ lattice artifacts, see \cite{Hasenfratz:1983ba}.
The fact that a staggered fermion represents four physical fermions is handled through the fourth-root trick and leads to a pressure for one physical fermion of
\begin{align}
\label{Pfullexpression}
    P & = 2 
    \int_{-\pi/2a}^{\pi/2a} \frac{d^3 \vec p}{(2\pi)^3}
    \int_{-\pi/2a}^{\pi/2a} \frac{dp_0}{2\pi}
    \left( \ln \frac{\sin^2(a p_0+ia \mu) + a^2 E_p^2}
    {\sin^2(ap_0) + a^2 E_p^2} \right)
    \nonumber \\
    a^2 E_p^2 & = \sum_j \sin^2(a p_j) + a^2 m^2 \,.
\end{align}
There are two effects.
First, the lattice, rather than physical, dispersion determines the energy $E_p$.
Second, the $p_0$ range is finite and periodic and $p_0$ appears inside a trigonometric function inside the log.
Nevertheless, a contour deformation, $p_0 \to p_0 - i\mu$, is still possible, and it encounters a similar discontinuity in the log:
\begin{align}
\label{Pnaivequarks}
    P & = 2 \int_{-\pi/2a}^{\pi/2a} \frac{d^3 \vec p}{(2\pi)^3}
    \left( |\mu| - a^{-1} \mathrm{arcsinh} (aE_p) \right)
    \Theta \left(|\mu| - a^{-1} \mathrm{arcsinh} (a E_p) \right) \,.
\end{align}
We have only been able to compute the resulting $\vec p$ integral numerically, but as expected, the leading small-$\mu$ corrections are of order $\mu^2 a^2$.
Specifically, the flatter dispersion relation lowers $E_p$ and leads to a larger phase space region where fermionic states are occupied.

\begin{figure}[htb]
    \centerline{
    \includegraphics[width=0.9\textwidth]{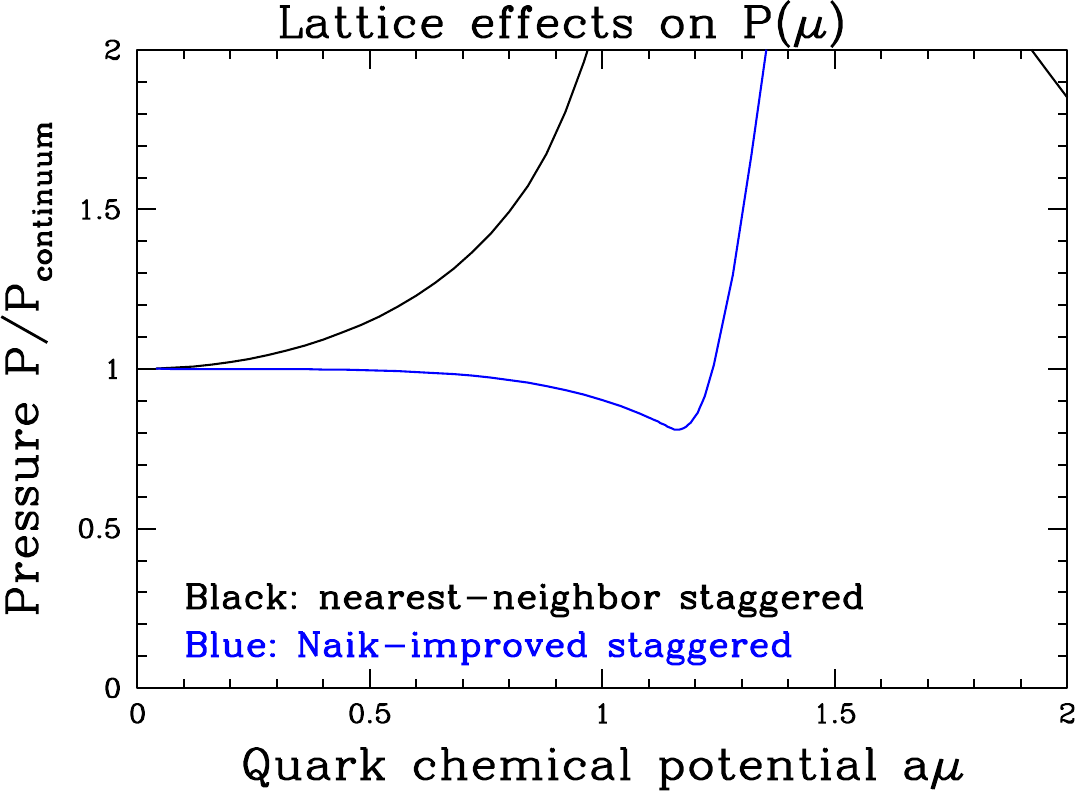}
    }
    \caption{Lattice-to-continuum pressure ratio as a function of the chemical potential in lattice units.
    The black curve is for nearest-neighbor staggered fermions, the blue curve is for 3'rd-neighbor-improved fermions.
    At the order of interest, the use of ``fattened links'' and other modifications to the fermionic action are not relevant.
    \label{fig:latt_mu}
    }
\end{figure}

We show the results of a simple numerical evaluation of \Eq{Pnaivequarks} in Figure \ref{fig:latt_mu}.
We have also checked that carrying out the integral shown in \Eq{Pfullexpression} leads to the same result.
Unfortunately, if we set as a criterion for good convergence that the lattice and continuum pressures should differ by at most 10\%, we are restricted to $a\mu < 0.42$.
To do better, we could use an improved fermionic action.
Naik has advocated
\cite{Naik:1986bn}
that one replace \Eq{nearestneighbor} with an improved version,
\begin{align}
\label{Naik}
    \partial_0 \psi(x) + \mu \psi(x) & \quad \Longrightarrow \quad
    \frac{9}{8} \frac{e^{a\mu} \psi(x{+}a) - e^{-a\mu} \psi(x{-}a)}{2a}
    - \frac{1}{8} \frac{e^{3a\mu} \psi(x{+}3a\hat{0})
    - e^{-3a\mu} \psi(x{-}3a \hat{0})}{6a}
    \nonumber \\
    \sin^2(ap_j) & \quad \Longrightarrow \quad
    \left( \frac{9}{8} \sin(ap_j) - \frac{1}{24} \sin(3ap_j) \right)^2\,,
    \nonumber \\
    \sin^2(ap_0+ia\mu) & \quad \Longrightarrow \quad
    \left( \frac 98 \sin(ap_0+ia\mu) - \frac{1}{24} \sin(3ap_0+3ia\mu) \right)^2 \,.
\end{align}
As shown in Figure \ref{fig:latt_mu}, this dramatically improves the match between the lattice and continuum pressure, such that $a\mu=1$ still has $<10\%$ corrections.
However, beyond $a\mu = 1.147$, $\cosh(3a\mu) > 9 \cosh(a\mu)$ and the ``improvement'' term in \Eq{Naik} starts to dominate over the nearest-neighbor term.
This leads to additional cuts in the modified version of \Eq{Pfullexpression}, and the performance of this implementation rather abruptly breaks down.

With these results in mind, we feel that nearest-neighbor staggered quarks can only treat large $\mu$ accurately out to disappointingly small $\mu a \sim 0.4$ or 0.5 -- possibly a little higher with the help of extrapolation over a few lattice spacings.
Improved quarks, such as ASQTAD
\cite{Orginos:1998ue,Lepage:1998vj}
or HISQ
\cite{Follana:2006rc}
quarks, which use the Naik term,
should be able to do better but it is very dangerous to venture beyond $a\mu = 1.1$.

Another limitation of a lattice treatment is that one generically must compute at finite temperature.
Therefore we should also estimate the size of thermal effects.
We will assume that the thermal corrections are similar to those in the continuum.
At the free theory level, the pressure of a single Dirac fermion with chemical potential $\mu$ at temperature $T$ actually has a closed form:
\begin{equation}
\label{PvsT}
    P = 4 \left( \frac{7 \pi^2}{720} T^4
    + \frac{1}{24} \mu^2 T^2 + \frac{1}{48\pi^2} \mu^4 \right).
\end{equation}
This implies $\mu > 14 T$ to keep thermal effects at the 10\% level.

One can attempt to remove both lattice-spacing and temperature effects through extrapolation over multiple lattice spacings and box sizes.
In the case of lattice spacing effects, an effective field theory analysis at the scale $\mu$ tells us that the lattice-spacing effects should scale as $(\mu a)^2$ up to anomalous dimension corrections.
The coefficient will not necessarily equal the free-theory one, but the effect should definitely be a power law with power close to 2.
More care is needed in extrapolating away temperature effects.
While one might expect that the leading thermal effects are of order $T^2$ as in \Eq{PvsT}, this is really an assumption which can go wrong if, for instance, the theory develops a mass gap due to interactions between excitations near the Fermi surface.
Therefore we expect that more care must be taken when extrapolating to small temperature.

\section{Discussion}

This paper has made three points.
\begin{itemize}
\item 
    Lattice QCD can study arbitrary combinations of $\mu_u,\mu_d,\mu_s$ using \textsl{phase-quenching}, providing $\PPQ(\mu_q)$.
    Though this does not return the true QCD pressure, it returns a \textsl{strict upper bound} on the true pressure: $\PPQ(\mu_q) \geq P(\mu_q)$, which could still be useful in constraining the equation of state for neutron star matter.
    \item 
    The difference $\PPQ-P$ is small at weak coupling, in the sense that $\frac{\PPQ-P}{P} \propto \alphas^3$.
    Furthermore, the $\alphas^3$ contribution arises from a \textsl{single} diagram.
    If we could compute this diagram, we could use $\PPQ$ to determine $P$ up to $\alphas^4$ corrections (up to logs and nonperturbative effects such as pairing gaps).
    \item 
    Lattice calculations of $P(\mu)$ encounter lattice artifacts.
    We estimate the size of these artifacts and advocate that nearest-neighbor fermion formulations use $a \mu \leq 0.4$, while improved-dispersion fermions may be reliable at chemical potentials more than a factor of 2 larger.
\end{itemize}

The first step in utilizing this approach is to perform a lattice study at a series of chemical potentials.
For neutron star physics one should choose $\mu_d = \mu_s$ and $\mu_u$ somewhat smaller to describe a charge-neutral system including equilibrated leptons.
One might study a range of chemical potentials from $\mu_s = 100$ MeV to 1 GeV.
Since a lattice approach will likely involve determining $N$ at each $\mu$ value and integrating it to find the pressure using $N_q = dP/d\mu_q$, it appears necessary to consider a tightly spaced series of $\mu$ values.
One also needs a few lattice spacings in order to perform a continuum extrapolation.

The next step would be to use these $\PPQ(\mu)$ values as constraints, when considering the high-density QCD equation of state.
A proposed QCD equation of state is usually expressed as a curve in either the $(P,N)$, $(N,\varepsilon)$, or $(P,\varepsilon)$ plane.
Standard thermodynamical relations can convert this into a curve in the $P,\mu$ plane; if the proposed EOS exceeds $\PPQ(\mu)$ for any $\mu$ where we have data, it is excluded.

Next, the possibility of evaluating $\PPQ(\mu)$ directly on the lattice at large $\mu$ can be used as a check on the performance of the perturbative expansion.
Specifically, one can consider the perturbative expansion for $\PPQ(\mu)$, which as we have argued is the same as the expansion for $P(\mu)$ at the currently known order of $\alphas^3 \ln(\alphas)$
\cite{Gorda:2023mkk}.
By comparing this result to the lattice-determined $\PPQ(\mu)$, one can assess how accurate the perturbative series is as a function of the scale $\mu$.

Finally, one should perform an accurate evaluation of the single diagram which introduces $\OO(\alphas^3)$ differences between $\PPQ$ and $P$.
In any $\mu$-region where the result is actually a small correction, this can be used together with a lattice-determined $\PPQ(\mu)$ to provide an improved estimate of $P(\mu)$.
The resulting estimate would also be perturbatively complete at $\OO(\alphas^3)$, with only $\alphas^4 \ln(\alphas)$, higher-order, and nonperturbative corrections remaining.

\section*{Acknowledgments}

We would like to thank Gergely Endr\H{o}di and Saga S\"appi for useful discussions.
We acknowledge support by the Deutsche Forschungsgemeinschaft (DFG, German Research Foundation) through the CRC-TR 211 ``Strong-interaction matter under extreme conditions''– project number 315477589 – TRR 211, the DFG-Project ID 279384907--SFB 1245, and by the State of Hesse within the Research Cluster ELEMENTS (Project ID 500/10.006).

\bibliographystyle{JHEP}
\bibliography{refs}

\end{document}